\documentclass[letterpaper]{article} 
\usepackage{aaai2026}  
\usepackage{times}  
\usepackage{helvet}  
\usepackage{courier}  
\usepackage[hyphens]{url}  
\usepackage{graphicx} 
\usepackage{natbib}  
\usepackage{caption} 
\usepackage{algorithm}
\usepackage{algorithmic}

\usepackage{newfloat}
\usepackage{listings}
\DeclareCaptionStyle{ruled}{labelfont=normalfont,labelsep=colon,strut=off} 
\floatstyle{ruled}
\newfloat{listing}{tb}{lst}{}
\floatname{listing}{Listing}
\title{InfoDecom: Decomposing Information for Defending Against Privacy Leakage in Split Inference}
\author{
    Ruijun Deng\textsuperscript{\rm 1},
    Zhihui Lu\textsuperscript{\rm 1}\thanks{Corresponding authors.},
    Qiang Duan\textsuperscript{\rm 2}
}
\affiliations{
    \textsuperscript{\rm 1}College of Computer Science and Artificial Intelligence, Fudan University\\
    \textsuperscript{\rm 2}Pennsylvania State University\\

    \{rjdeng18,lzh\}@fudan.edu.cn,qduan@psu.edu
}

\usepackage{bibentry}

\usepackage{multirow}
\usepackage{booktabs} 
\usepackage{amssymb}
\usepackage{amsmath}
\usepackage{mathtools}
\usepackage{amsthm}
\usepackage{graphicx}
\usepackage{subfigure}
\usepackage{makecell}

\begin{document}

\maketitle

\begin{abstract}
Split inference (SI) enables users to access deep learning (DL) services without directly transmitting raw data. However, recent studies reveal that data reconstruction attacks (DRAs) can recover the original inputs from the smashed data sent from the client to the server, leading to significant privacy leakage. While various defenses have been proposed, they often result in substantial utility degradation, particularly when the client-side model is shallow. We identify a key cause of this trade-off: existing defenses apply excessive perturbation to redundant information in the smashed data. To address this issue in computer vision tasks, we propose InfoDecom, a defense framework that first decomposes and removes redundant information and then injects noise calibrated to provide theoretically guaranteed privacy. Experiments demonstrate that InfoDecom achieves a superior utility-privacy trade-off compared to existing baselines.
\end{abstract}

\begin{links}
    \link{Code}{https://github.com/SASA-cloud/InfoDecom}
    \link{Extended version}{https://arxiv.org/abs/2511.13365}
\end{links}

\section{Introduction}

\begin{figure}[!t]
  \centering
  \includegraphics[width=1.0\linewidth]{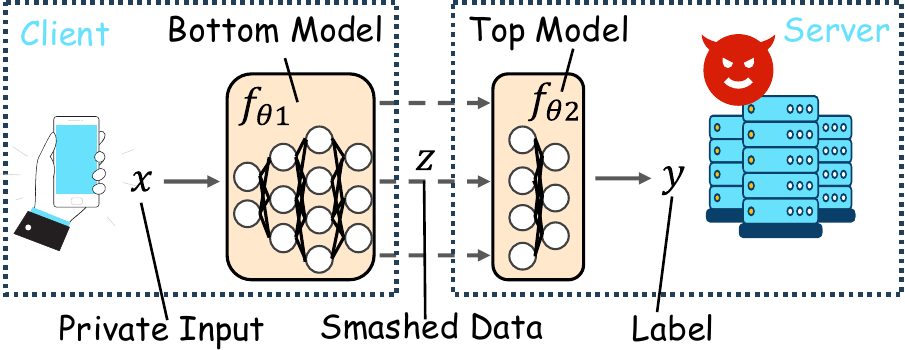}
  \caption{A general framework of two-party split inference.}
  \label{fig:SI-Scenario}
\end{figure}

Machine learning (ML) has achieved breakthroughs in many areas of computer vision. Given the rising size of deep learning (DL) models and substantial costs associated with model accommodation, it is challenging to implement DL inference applications on the resource-constrained user/edge devices (e.g., mobile phones or smart cameras). This has led to the emergence of ML as a service (MLaaS) \cite{ccs24-SL-LLM}, a paradigm that allows enterprises with sufficient resources to accept the inference request from user devices, execute on the server, and return the results. Split inference (SI) or collaborative inference \cite{nips24-club,deng2023hsfl,ijcai2024p596} is one widely employed way to achieve MLaaS, where a DL model is divided into two parts. Figure~\ref{fig:SI-Scenario} shows a two-party SI scenario, where the first part (bottom model) is shallow and deployed on the user device or the client, while the remaining layers (top model) are offloaded to the server. This division allows the client to submit only the output of the bottom model, i.e., the smashed data or the representation, to the server, instead of the raw input data, and is thus expected to preserve feature privacy.

However, recent works show that SI is susceptible to privacy attacks, e.g., the data reconstruction attacks (DRAs) \cite{blackbox,nips2023gan}. A malicious DRA adversary (e.g., the untrusted server) can reconstruct the raw input from the smashed data, posing a great threat to the user's privacy, as once successfully conducted, the whole dataset rather than specific private properties will leak.

To mitigate these privacy threats, defensive works mainly use perturbation-based methods to defend against inference attacks in SI \cite{singh2024simba,ccs24-SL-LLM}. These methods follow two lines: i) \textit{Regularization} \cite{arevalo2024task,aaai25-robust-pp,nips24-club}, which optimizes the model or the parameterized noise/mask to output perturbed smashed data with heuristic regularization terms; and ii) \textit{closed-form noise calculations} \cite{singh2023posthoc,deng2025quantifying,sec24-defending-FL-ita}, which rely on theories like metric Differential Privacy (metric-DP), 
Mutual Information (MI), 
Fisher Information (FI)
and conditional entropy, 
for obtaining analytical solutions of the perturbing noise scale that satisfies the desired security level. The regularization defenses have no stringent provable privacy guarantees, as they optimize the heuristic objectives (e.g., minimize the similarity between the raw input and the reconstructed one from the DRA adversary \cite{sun2021soteria}). Closed-form noise calculations solve this by establishing the analytical relationship between certain privacy metrics/budgets and the perturbation noise scale. 

However, when the bottom model is shallow, neither method can achieve a good utility-privacy trade-off (UPT), as shown in our results. Deeper bottom models inherently filter out task-irrelevant information through non-linear transformations, making them more privacy-resilient and easier to balance utility and privacy. In contrast, when the bottom model is shallow—a common scenario due to resource-constrained user devices \cite{deng2023hsfl}, it retains more input information in the smashed data, requiring heavy obfuscation to ensure privacy, which in turn degrades task performance. Hence, current defense mechanisms struggle to perform well under shallow client models, highlighting the need for more effective and adaptive privacy-preserving solutions in realistic deployment settings.

In this study, we propose InfoDecom, a defense method against DRAs to protect user data privacy in SI systems for visual tasks. Our goal is to provide strong, theoretically grounded privacy protection for shallow client models without significantly compromising task accuracy. We argue that the limited UPT in existing defenses stems from the need to protect excessive input information in the smashed data. However, much of this information is task-irrelevant or has minimal impact on the final prediction, and existing defenses waste perturbation on it. InfoDecom decomposes and filters out such redundant information, reducing the volume of sensitive content that needs protection. As a result, less noise is required to achieve a given theoretical privacy level, leading to reduced performance degradation and improved UPT. 

To generate privacy-preserving smashed data, InfoDecom incorporates three key components. First, input images are transformed into the frequency domain, retaining only channels that are non-essential for human perception. Second, the remaining information is further suppressed using regularization terms inspired by the Information Bottleneck (IB) principle. Third, we apply a closed-form calculation to determine the appropriate Gaussian noise scale, ensuring a guaranteed privacy level is met.
The main contributions of this study are summarized as follows:
\begin{itemize}
    \item \textbf{Guaranteed privacy with good utility-privacy trade-off}: InfoDecom considers the information redundancy in smashed data, achieving theoretical privacy guarantees against DRAs in shallow-client SI vision systems while maintaining high task utility.
    \item \textbf{Simple yet effective defense design}: InfoDecom adopts a two-stage approach to remove redundant information and applies noise perturbation to enforce a target privacy level.
    \item \textbf{Superior performance over state-of-the-art (SOTA) defenses}:
    Experiments on multiple vision benchmarks show that InfoDecom consistently outperforms existing methods in defending against DRAs while preserving model performance.
\end{itemize}

\section{Related Work}

\subsection{Regularization}
Defense methods falling into this type add regularization terms to loss functions during model training, guiding the (perturbed) smashed data to have the desired properties. Typically, these terms can be categorized into two goals: reducing privacy leakage (Goal 1) and maintaining task performance (Goal 2). A detailed summary of the regularization loss terms from previous works is presented in the Appendix.

\textbf{Heuristic strategies.} Some methods adopt heuristic optimization targets on smashed data. For example, Soteria \cite{sun2021soteria} achieves Goal 1 by minimizing the negative $L_p$ norm of raw input data $x$ and the reconstructed one $\hat{x}$, and Goal 2 by constraining the difference $L_q$ norm of original smashed data and the perturbed one. ML-ARL \cite{roy2019mitigating} formulates an adversarial representation learning problem that achieves Goal~1 by maximizing the uncertainty (entropy) of the adversary and achieves Goal~2 by minimizing the Kullback-Leibler (KL) divergence between the distribution of ground truth and the predicted one. Nopeek \cite{ICDM-20-nopeek} reduces the distance correlation between input data and smashed data as well as the normal classification cross-entropy loss.

\textbf{Mutual Information Optimization.} Mutual information (MI) is a widely adopted regularization guidance. Cloak \cite{www21-cloak} trains the noise mask on input data by suppressing the MI between the perturbed input and the redundant features while enhancing the MI between the perturbed input and the task-useful features. Shredder \cite{mireshghallah2020shredder} minimizes the MI between the smashed data and the raw input (with a signal-to-noise ratio approximation) as well as the cross-entropy loss. Inf2Guard \cite{noorbakhsh2024inf2guard} minimizes similar MI terms with Shredder but uses Jensen-Shannon divergence (JSD) to approximate the MI. TAPPFL \cite{arevalo2024task} adopts the same MI approximation methods as Inf2Guard, but it focuses on the MI between the private attribute instead of the raw input and the smashed data. DPFE \cite{tkde20-dpfe} minimizes the MI between smashed data and the private feature while maximizing the MI between smashed data and the task label. It employs the kernel density estimation for MI approximation. InfoScissors \cite{nips24-club} restricts the MI between smashed data and the input by minimizing the upper bound (CLUB \cite{cheng2020club}) of the MI. ARPRL \cite{aaai25-robust-pp} has the same goal as TAPPFL.

\subsection{Closed-Form Noise Calculation}
All the aforementioned studies only rely on optimizations to get perturbed smashed data. However, these empirical approaches forgo theoretical guarantees of bounding privacy leakage. The provable defenses, e.g., DP-SGD \cite{abadi2016deep}, are predominantly centered around differential privacy (DP) \cite{dwork2006differential} for theoretically bounding the adversary's reconstruction error and protecting data identity \cite{singh2024simba,sec24-defending-FL-ita}. However, the vanilla DP is not applicable for the SI system, as it requires that outputs of any two samples are indistinguishable \cite{singh2023posthoc,deng2025quantifying}. Therefore, other theoretical privacy metrics (e.g., Fisher information \cite{martens2020new}) are derived for analyzing the robustness of SI. Just like the privacy budget $\epsilon$ in DP, privacy metrics quantitatively measure the privacy leakage level of a system and are proven to bound the adversaries' error.

Given a privacy metric or budget, the corresponding closed-form noise scale (e.g., the standard deviation of a Gaussian distribution) can be directly computed based on model analysis over the given dataset \cite{maeng2023bounding,deng2025quantifying,singh2023posthoc,sec24-defending-FL-ita}. For example, dFIL \cite{maeng2023bounding} is a privacy metric that relies on Fisher information about raw input in smashed data. According to its definition, the scale of the Gaussian distribution that needs to be added to achieve a certain dFIL is 
\begin{equation}
    \sigma=\sqrt{\frac{\text{Tr}(J^{\top}_{f_{\theta 1}} J_{f_{\theta 1}})}{d \times \text{dFIL}}},
\end{equation}
where $f_{\theta 1}$ is the bottom model and $x$ is the input with dimensions of $d$.
Similarly, the noise scale to achieve a certain privacy metric FSInfo \cite{deng2025quantifying} is
\begin{equation}
    \sigma=\frac{det(J^{\top}_{f_{\theta 1}} J_{f_{\theta 1}}))^\frac{1}{2d}}{e^{\text{FSInfo}}(2\pi e)^\frac{1}{2}}.
\end{equation}

However, when the client-side bottom model is shallow, leaving substantial information in smashed data, these defenses tend to obtain a large noise scale from the desired privacy leakage level or metrics. Therefore, they encounter substantial model performance degradation to achieve a decent defense performance. To address the aforementioned issues, we propose InfoDecom by decomposing the utility- and privacy-related information to gain a decent and theoretically guaranteed privacy robustness without huge performance degradation on the shallow bottom model.

\begin{figure}[!t]
  \centering
  \includegraphics[width=0.7\linewidth]{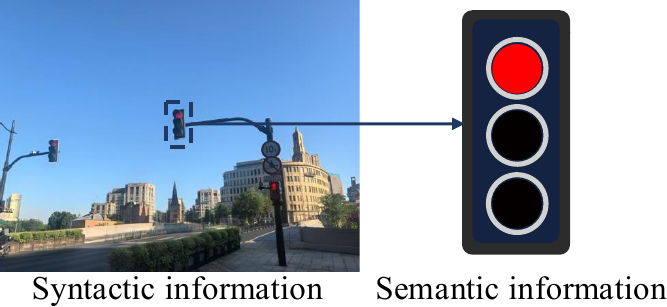}
  \caption{Example of semantic-oriented communication.}
  \label{fig:semantic_info}
\end{figure}

\section{Motivations}
\subsection{Triple Definitions of Communication}

DNN inference is an information processing procedure where data communications occur between adjacent layers.
Semiotics of communication can be defined as a triple combination of syntactics, semantics, and pragmatics \cite{yang2022semantic}. \textit{Syntactics} focuses on the communication symbols or formal features of signs (visual and linguistic), ensuring that each data bit is transmitted. \textit{Semantics} specializes in the meaning of the signs. \textit{Pragmatics} concentrates on the contribution of communicated information to the task.

\subsection{Redundancy in Syntactic Communication}
Syntactic communication—such as feeding raw images directly into a DNN—often transmits excessive redundant information that is eventually discarded during inference. In contrast, semantic-oriented communication focuses on conveying only the task-relevant meaning. As illustrated in Figure~\ref{fig:semantic_info}, for tasks like traffic signal recognition, transmitting only the signal light rather than the full image is sufficient.
Moreover, SCA \cite{eccv24-SCA} examines model architectures and shows that incorporating a sparse coding layer—which filters out task-irrelevant details and retains only essential information—enhances robustness against DRAs while maintaining model accuracy.
These observations suggest that \textit{reducing input-space redundancy can effectively mitigate privacy leakage from DRAs with minimal impact on task performance.}

\subsection{Redundancy in Semantic Communication}
While semantic-oriented communication removes irrelevant input details, it remains unclear how effectively the retained meaning contributes to the task objective. The Information Bottleneck (IB) framework \cite{tishby2000information} addresses this by formulating the problem as:
\begin{equation}
     \min_Z \lambda I(X;Z) - I(Y;Z), \label{eq:ib-def}
\end{equation}
where $X$, $Y$, and $Z$ denote the input, output, and intermediate representation (i.e., smashed data) of a DNN, respectively. $I(\cdot;\cdot)$ denotes mutual information \cite{cover1999elements}, and $\lambda > 0$ balances compression and relevance. The IB principle aims to learn a minimal yet sufficient representation of the input that preserves task-relevant (i.e., pragmatic) information, thereby reducing communication overhead compared to syntactic or purely semantic approaches.
This principle provides \textit{a theoretical foundation for guiding the client-side model to produce smashed data that retains task-relevant information while suppressing task-irrelevant content.}

\section{Design of InfoDecom}
In this section, we introduce InfoDecom, a novel framework for privacy-preserving split inference that decomposes and reduces redundant and private information.

\subsection{Overview}

\begin{figure*}[!th]
  \centering
  \includegraphics[width=0.9\linewidth]{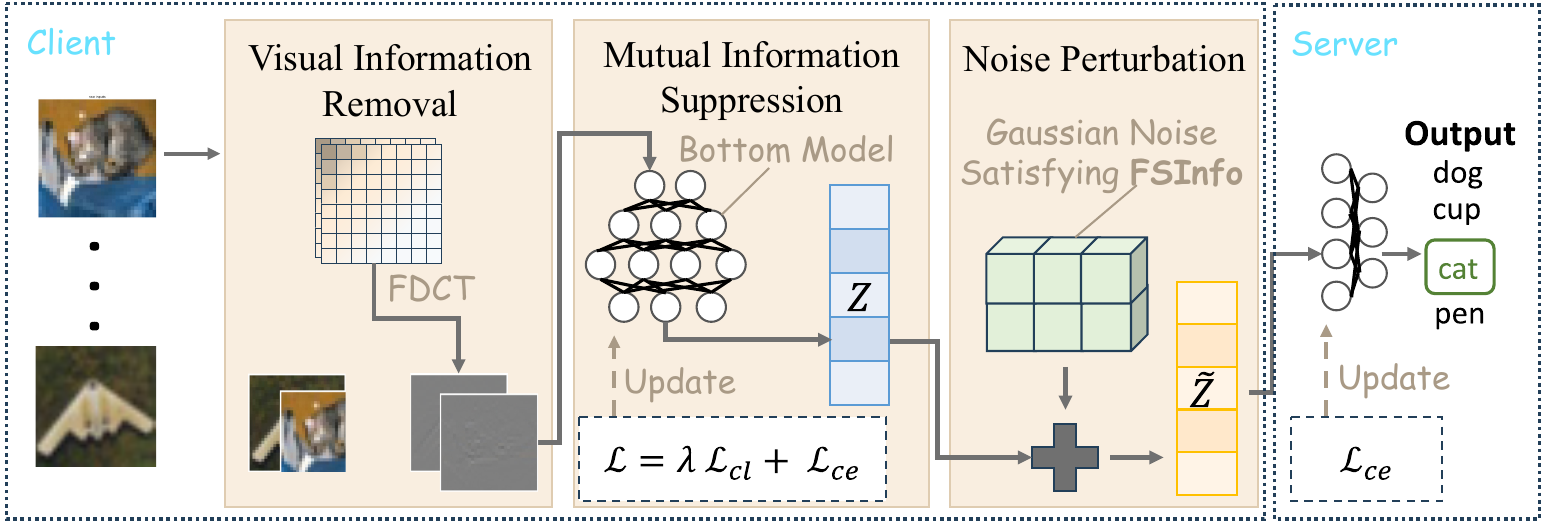}
  \caption{The overview of InfoDecom.}
  \label{fig:InfoDecom-Overview}
\end{figure*}

\textbf{SI System.}
W.l.o.g., Figure~\ref{fig:SI-Scenario} illustrates a two-party SI system comprising a client $C$ with private data $D$ and a server $S$. The deep learning model $f_\theta(\cdot) = (f_{\theta2} \circ f_{\theta1})(\cdot) = f_{\theta2}(f_{\theta1}(\cdot))$ is partitioned into a bottom model $f_{\theta1}$ deployed on $C$ and a top model $f_{\theta2}$ on $S$. The split point (SP) $p \in {1, 2, \dots, L}$ denotes the last layer of $f_{\theta1}$, where $L$ is the total number of layers. During inference, $C$ takes input $x$ and sends the smashed data $z = f_{\theta1}(x)$ to $S$, which completes the inference by computing $y = f_{\theta2}(z)$. Notably, the bottom model $f_{\theta1}$ is usually shallow on the resource-constrained client devices.

\textbf{Threat Model.} 
In this paper, both $C$ and $S$ are assumed to strictly follow the learning protocol. However, the server is considered honest-but-curious, aiming to infer private information from the received intermediate representations $z$. We focus on the DRA setting, where the untrusted server attempts to reconstruct the original input $x$ via an attack function $g$: $\hat{x} = g_{\phi}(z)$, with $\hat{x}$ denoting the reconstructed input. In the vision domain, we treat the visual recognizability of $x$ as the private information rather than a certain attribute.

We follow the assumptions of the SOTA DRAs \cite{nips2023gan,yin2023ginver,blackbox} about the adversaries' knowledge and capabilities. The adversary $\mathcal{A}$ is allowed to have auxiliary information, such as a surrogate bottom model (including parameters and the architecture), a dataset with a similar distribution to $D$, and some prior information about $D$. The adversary can alleviate the optimization-based, learning-based, or combined DRA methods to reconstruct the raw input after seeing the smashed data $z$. However, it cannot interfere with the normal inference process. 

\textbf{Workflow.} Our approach is based on the observation of the redundancy of information flow in the DNN. 
Figure~\ref{fig:InfoDecom-Overview} illustrates the InfoDecom workflow: a two-stage information elimination---targeting visual and mutual information---followed by closed-form noise perturbation. First, the input is decomposed in the frequency domain into essential (low-frequency) and non-essential (high-frequency) components for visual perception, and the essential (private) channels are discarded. Next, based on the IB principle, the smashed data is further decomposed into task-relevant and task-irrelevant information. The bottom model is guided to retain the former while suppressing the latter. Finally, closed-form Gaussian noise is added to the smashed data to provide theoretical privacy guarantees. During inference, high-frequency-only inputs are passed through the updated bottom model to produce regularized smashed data, which is then perturbed before transmission to the server.

\subsection{Visual Information Removal}
\begin{figure}[!t]
  \centering
  \includegraphics[width=1.0\linewidth]{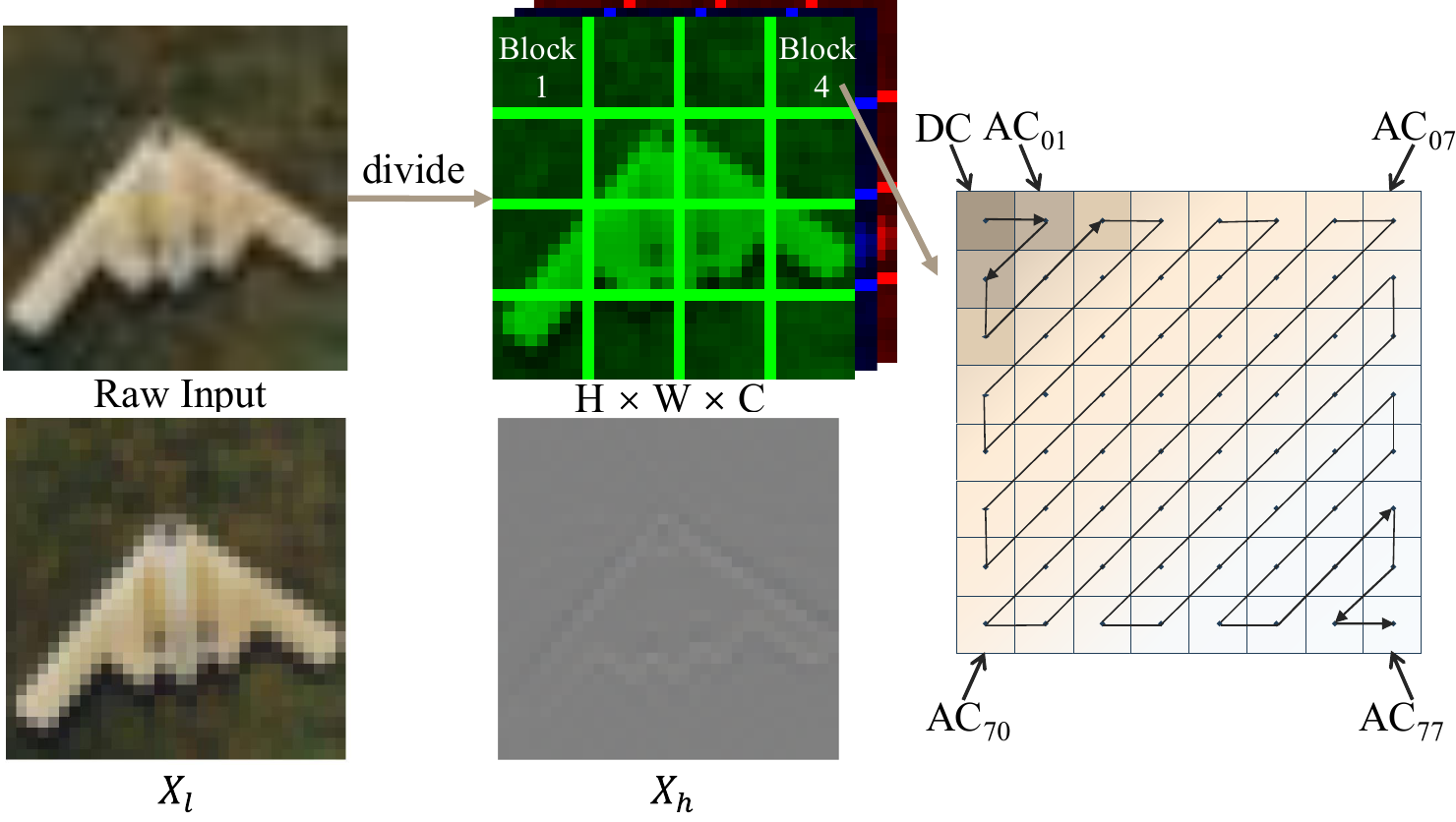}
  \caption{Visual information removal for raw input. The raw input is divided into blocks, with each block being transformed to 8$\times$8 DCT coefficients.}
  \label{fig:visual-fdct}
\end{figure}
To perform image classification, DNNs require access to private inputs. Most SI systems transmit the entire image to the DNN, forming a syntactic communication pattern. However, as previously discussed, syntactic communication often contains redundancy. For instance, prior work~\cite{jin2024faceobfuscator} shows that many frequency channels can be removed with minimal impact on face recognition accuracy.
Processing images in the frequency domain has long been used for image compression, retaining low-frequency components crucial for human visual perception while discarding high-frequency details (e.g., subtle textures) that are less noticeable~\cite{wallace1991jpeg}. However, DuetFace~\cite{mi2022duetface} demonstrates that high-frequency components still carry sufficient semantic information for DNNs to complete classification tasks.
Motivated by this, we initiate our information decomposition from the visual dimension---removing private, visually sensitive low-frequency channels and retaining less private, high-frequency ones in the frequency-domain representation.

To transform the images to the frequency domain, we follow the building blocks in standard JPEG compression \cite{wallace1991jpeg}. First, we transform the image from RGB to YUV color space. Then, each component (e.g., the two-dimensional luma Y) is further divided into $8\times8$ blocks, which will be grouped as the input for the Forward Discrete Cosine Transform (FDCT) process. The FDCT decomposes each $8\times 8$ block into 64 orthogonal frequency signals or ``DCT coefficients'', with each capturing one frequency of the $8\times 8$ spatial signals. As shown in Figure~\ref{fig:visual-fdct}, the DCT coefficients contain one ``DC coefficient'' and 63 ``AC coefficients''. When ordered into the ``zig-zag'' sequence, the more likely non-zero low-frequency coefficients are placed before the near-zero high-frequency coefficients.

Following DuetFace \cite{mi2022duetface}, we delete the $K$ DCT coefficients with the highest amplitude, i.e., the low-frequency coefficients $X_l$, and only reserve the high-frequency coefficients $X_h$ for the DNN to perform task learning, as illustrated in Figure~\ref{fig:visual-fdct}. Therefore, most visual information is discarded, and sufficient semantic information is obtained.

\subsection{Mutual Information Suppression}

Although some visual details have been removed, the remaining frequency components $X_h$ may still contain privacy-sensitive information exploitable by DRA adversaries. To this end, we perform decomposition in the mutual information plane and introduce regularization terms that encourage the bottom model to extract pragmatic representations, i.e., those informative for the task, while suppressing task-irrelevant private information before feeding it into the top model.

\textbf{Formal Goals.}
We use the IB principles (see Equation~\ref{eq:ib-def}) to remove the information redundancy in smashed data. The optimization problem is formulated as:
\begin{equation}
    \min_{Z}\quad \lambda I(X_h;Z)-I(Y;Z),
\end{equation}
where the $I(X_h;Z)$ represents the information of the high-frequency coefficients $X_h$ contained in $Z$, the more of which indicates higher privacy leakage. Similarly, the $I(Y;Z)$ represents the contribution of $Z$ to the task goal $Y$. $\lambda$ is a hyperparameter controlling the trade-offs.

\textbf{Minimize $I(X_h;Z)$.} Calculating the closed-form MI of high-dimensional random variables with arbitrary distributions is still an open problem \cite{deng2025quantifying,nips24-club,cheng2020club}. There are attempts for lower- and upper-bound MI estimation. CLUB \cite{cheng2020club} is an MI upper bound, where MI is estimated by the difference between positive and negative sample pairs (in a contrastive learning manner). The MI upper bound minimization with the variational version of CLUB (vCLUB) is:
\begin{equation}
    \hat{I}_{vCLUB}(X;Z) = \frac{1}{N}\sum_{i=1}^N \log q_1(z_i|x_i)-\log q_1(z_{j}|x_i), 
\end{equation}
where $q_1(Z|X)$ is a variational approximation of true conditional distribution $p(Z|X)$, $N$ is the dataset size, and $j$ is uniformly sampled from $\{1,2,\dots,N\}$.

In most machine learning tasks, both conditional distributions ($p(Z|X)$ and $q_1(Z|X)$) are inaccessible. Previous works using the reparameterization trick \cite{kingma2013auto}, training a neural network to approximate $q_1(z|x)$. However, this needs a laborious training stage, and the estimation may have a large bias due to suboptimal hyperparameters. To avoid this, we use CLUB as a guidance for designing a tractable loss term. The CLUB loss penalizes the model for assigning a high conditional likelihood to the true smashed data $z_i$ given input $x_i$, while encouraging higher conditional likelihoods for mismatched pairs $(x_i,z_i),j\neq i$, thereby increasing ambiguity in the conditional distribution and impeding adversarial inversion from latent representations back to original inputs.

Following this, we design the clustering loss $\mathcal{L}_{cl}$ to push the smashed data $z_i$ of different input $x_i$ to be entangled or to have smaller pairwise distance, thereby reducing their distinguishability and enhancing privacy:
\begin{equation}
    \mathcal{L}_{cl} = \frac{1}{N}\sum_{i=1}^N \Vert z_i - z_j\Vert_2^2 ,\label{eq:lcl}
\end{equation}
where $\Vert \cdot \Vert_2$ is the L2-norm, $z$ is the smashed data and index $j$ is uniformly sampled from $\{1,2,\dots,N\}$.

\textbf{Minimize $-I(Y;Z)$.} 
Barber-Agakov (BA) lower bound \cite{barber2004algorithm} of $I(Y;Z)$ is:
\begin{equation}
    \hat{I}_{BA}(Y;Z)=H(Y) + \frac{1}{N} \sum_{i=1}^N \log q_2(y_i|z_i), \label{eq:iyz}
\end{equation}
where $q_2(Y|Z)$ is a variational approximation of the true conditional distribution $p(Y|Z)$.

$H(Y)$ is a constant, and we omit this term during maximization. Similarly, to avoid the variational distribution ($q_2(Y|Z)$) approximation, we follow \cite{www21-cloak,aaai25-robust-pp,noorbakhsh2024inf2guard} to replace the second term of Equation~\ref{eq:iyz} with the negative empirical cross-entropy loss. To minimize $-I(Y;Z)$, we optimize the following loss:
\begin{equation}
    \mathcal{L}_{ce} = - \frac{1}{N}\sum_{i=1}^N \sum_{k=1}^K y_i^{(k)} \log(f_{\theta 2}(z_i))^{(k)}, \label{eq:lce}
\end{equation}
which is the expected cross-entropy loss, with the number of classes denoted by $K$.

\textbf{Noise Perturbation for Theoretical Guarantees.} 
The two-stage information decomposition extracts the pragmatic information while reducing privacy leakage. However, these mechanisms alone do not provide theoretical guarantees on the privacy robustness of SI systems. FSInfo \cite{deng2025quantifying}, a state-of-the-art privacy metric, quantifies the leakage in SI systems and offers a provable lower bound on the adversarial reconstruction error in DRA settings. A lower FSInfo means less privacy leakage. We leverage FSInfo to derive the scale of Gaussian noise applied to the smashed data:

\begin{equation}
    \tilde{Z} = Z + \delta, \label{eq:perturb}
\end{equation}
where $\delta \sim \mathcal{N}\left(0,\frac{\det(J^TJ)^{\frac{1}{2d}}}{e^{FSInfo}(2\pi e)^{\frac{1}{2}}}\right)$, and $J$ is the Jacobian of $Z$ with respect to the raw input $X$. This provides a closed-form solution for the noise required to achieve a target FSInfo level (e.g., $-1$), ensuring a quantifiable privacy guarantee.

Note that, without the proposed two-level information decomposition, adding noise directly to the vanilla smashed data can also achieve the same level of $FSInfo$. However, this will cause a huge performance degradation as the calculated noise scale of $\delta$ is large for protecting all the task-irrelated information, as shown in experimental results.

\textbf{Overall Loss.} 
By combining Equations~\ref{eq:lcl},\ref{eq:lce}-\ref{eq:perturb}, the total loss $\mathcal{L}$ becomes:
\begin{equation}
    \mathcal{L} = \lambda \mathcal{L}_{cl} +\mathcal{L}_{ce},
\end{equation}
where $\lambda$ is the weighting factor. During the training phase, the top model is optimized by $\mathcal{L}_{ce}$, and the bottom model is optimized by both $\lambda \mathcal{L}_{cl}$ and $\mathcal{L}_{ce}$.

\section{Experiments}
\subsection{Experiment Implementation}
\textbf{Datasets and Models.} 
We evaluated on CIFAR-10 \cite{krizhevsky2009learning} and CelebA \cite{liu2015faceattributes}. For CIFAR-10, we do the traditional decuplet-classification task. For CelebA, we built the binary attractiveness classification task following \cite{nips2023gan}.
We use ResNet-18 \cite{he2016deep} as the backbone model for both CIFAR-10 and CelebA.
deployed on the client side. 
Table~\ref{tab:models} shows the critical layers and the default SP for the adopted model. 

\begin{table}[!t]

    \centering
    \begin{tabular}{cc}
    \toprule
    Model  & Architecture \\
    \midrule
    \multirow{2}{*}{ResNet-18}&C64 (default)-AP-[BB64]*2-[BB128]*2 - \\
    &[BB256]*2-[BB512]*2-AAP-FC10\\
    \bottomrule
    \end{tabular}
    \caption{Statistics of ResNet-18. Convolutional layers are denoted by C, followed by the number of filters;  AveragePooling and Adaptive AveragePooling layers are AP and AAP; Fully Connected layer is FC with the number of neurons; Basic Block of ResNet is BB with the channel size.} 
    \label{tab:models}

\end{table}

\textbf{Attack Methods.} Existing DRAs can be categorized into DL-based \cite{blackbox,yin2023ginver,nips2023gan} and regularized maximum likelihood-based (rMLE) \cite{blackbox}. Empirical evidence demonstrates superior reconstruction quality of NN-based methods compared to MLE-based approaches \cite{yin2023ginver}. Therefore, we use the DL-based methods invNet \cite{blackbox} 
to reconstruct the input data. More details can be found in the Appendix.

\textbf{Baselines.} We compare InfoDecom with four baselines: i) inv\_dFIL\_def \cite{maeng2023bounding}, ii) Nopeek \cite{ICDM-20-nopeek}, iii) Shredder \cite{mireshghallah2020shredder}, and iv) FSInfoGuard \cite{deng2025quantifying}. The details of these defense methods can be found in the Appendix.

\textbf{Evaluation Metrics.} We follow the previous common settings for choosing metrics. For evaluating the model utility, we use the classification accuracy (\textbf{Acc.}) on test data as the metric. For evaluating the privacy robustness, we use the mean squared error (\textbf{MSE}), 
between the reconstructed data and the raw input and the visual invertibility as the metric, where $MSE(x,\hat{x})=\frac{1}{d}\sum_i^d(x_i-\hat{x}_i)^2$ and visual invertibility is the similarity of raw and reconstructed inputs by human perception \cite{sun2024privacy}. A higher MSE and lower visual invertibility mean better privacy protection.

\textbf{Hyperparameters.} Following common pre-process \cite{jin2024faceobfuscator}, the CelebA images are rescaled to $112\times 112$. Then, both images (CelebA and CIFAR-10) are rescaled to [0,1]. Finally, we normalize them with a variance and mean of 0.5, adjusting their range of values to [-1,1]. We apply Adam with a learning rate of $3\times e^{-4}$ and a weight decay of 0.01. The number of global training epochs $E$ is 150. The batch size $B$ is set to 128. All of the defensive methods are implemented with Pytorch 2.4.1 and Python 3.10. The experiments for CIFAR-10 are performed on a server equipped with two NVIDIA GeForce RTX 4090 24GB GPUs, and the experiments for CelebA are performed on four NVIDIA A100 80GB GPUs.

\subsection{The Effect of Information Controller}
\begin{table}[!t]
  \centering
  \begin{tabular}{ccc|cc}
    \toprule
    \multicolumn{3}{c|}{CIFAR-10} & \multicolumn{2}{c}{CelebA} \\
    \cmidrule(lr){1-3} \cmidrule(lr){4-5}
    $|X_h|$ & Acc. & MSE & Acc. & MSE \\
    \midrule
    54 & \textbf{0.7329}  & 0.0843  & \textbf{0.9693} & 0.1984 \\
    41 & 0.6905  & 0.1497  & 0.8036 & 0.3273 \\
    32 & 0.3645  & 0.2337  & 0.6135 & \textbf{1.1024} \\
    18 & 0.1004  & \textbf{0.2492}  & 0.6135 & 1.1022 \\
    \bottomrule
  \end{tabular}
  \caption{The effects of the number of retained coefficients $|X_h|$ in visual information removal.}
  \label{tab:effect-channel}
\end{table}
\begin{table}[!t]
  \centering
  \begin{tabular}{ccc|cc}
    \toprule
    \multicolumn{3}{c|}{CIFAR-10} & \multicolumn{2}{c}{CelebA} \\
    \midrule
    $\lambda$ & Acc. & MSE & Acc. & MSE \\
    \midrule
    1 &  \textbf{0.7570}  & 0.0822   & 0.9515 & 0.1925 \\
    10 & 0.7329  & 0.0843   & \textbf{0.9693} & 0.1942 \\
    20 & 0.7250  & 0.0854   & 0.8997 & 0.1950 \\
    60 & 0.5964  & \textbf{0.0887}   & 0.8673 & \textbf{0.2128} \\
    \bottomrule
  \end{tabular}
  \caption{The effects of the number of weighing factor $\lambda$ for $\mathcal{L}$ in mutual information suppression.}
  \label{tab:effect-lambda}
\end{table}
\begin{table}[!t]
  \centering
  \begin{tabular}{ccc|cc}
    \toprule
    \multicolumn{3}{c|}{CIFAR-10} & \multicolumn{2}{c}{CelebA} \\
    \cmidrule(lr){1-3} \cmidrule(lr){4-5}
    FSInfo & Acc. & MSE & Acc. & MSE \\
    \midrule
    -0.5& \textbf{0.7356}  & 0.0837  & 0.9298 & 0.1741 \\
    -1 &  0.7329  & 0.0843  & \textbf{0.9693} & 0.1942 \\
    -1.5& 0.7162  & 0.0874  & 0.8823 & 0.2197 \\
    -2 &  0.6874  & \textbf{0.0878}  & 0.8371 & \textbf{0.2450} \\
    \bottomrule
  \end{tabular}
  \caption{The effects of the number of the desired privacy leakage level $FSInfo$ for closed-form noise perturbation.}
  \label{tab:effect-fsinfo}
\end{table}

Three controllers in InfoDecom adjust the amount of information transmitted to the server. In this section, we investigate the impact of them on the utility–privacy tradeoff of InfoDecom by varying their values.

\textbf{Impact of the Number of Retained Coefficients $|X_h|$.} In the visual information removal stage, we retain only the high-frequency coefficients $X_h$ (out of 64 total) for bottom model inference, where a larger $|X_h|$ indicates more visually relevant information is preserved. As shown in Table~\ref{tab:effect-channel}, with $FSInfo$ set to $-1$ and $\lambda$ to $10$, reducing $|X_h|$ limits the visual information available to the bottom model. Overall, \textbf{as $|X_h|$ decreases, task accuracy drops while MSE increases across both datasets.} This indicates that retaining more input information improves utility but also makes it harder to obscure private information. Additionally, on the CelebA dataset, when $|X_h|$ is reduced to 32 or 18, the input retains too little visual information for the binary classifier to perform effectively, resulting in near-random predictions. Meanwhile, the images reconstructed by the attacker become nearly unrecognizable—often appearing as entirely red or blue, with no discernible features.

\textbf{Impact of the Weighting Factor $\lambda$.} In mutual information suppression, a higher $\lambda$ means letting the loss term that aims to reduce $I(X_h;Z)$ dominate the update of the bottom model, thus decreasing the transmitted information about $X_h$. 
As shown in Table~\ref{tab:effect-lambda}, where $|X_h|$ is fixed at 54 and $FSInfo$ is set to $-1$, \textbf{increasing $\lambda$ leads to a decline in utility but enhances privacy protection.} However, on the CelebA dataset, we observe that the accuracy at $\lambda=1$ is slightly lower than at $\lambda=10$. This counterintuitive result stems from the fact that, to meet the same privacy guarantee ($FSInfo = -1$), more noise is required when $\lambda$ is smaller, which in turn degrades model performance.

\textbf{Impact of the Privacy Leakage Level $FSInfo$.} $FSInfo$ specifies the privacy leakage level of the SI system. To achieve less privacy leakage or a smaller $FSInfo$, a larger scale of Gaussian noise is needed. As shown in Table~\ref{tab:effect-fsinfo}, where the $|X_h|$ is set to 54 and the $\lambda$ is set to 10. We observe that \textbf{decreasing $FSInfo$ or the target privacy leakage impairs the utility but improves the privacy protection}. Similarly, there is an exception when the $FSInfo=-0.5$ on CelebA has similar but lower accuracy compared with that when $FSInfo=-1$. This may be because the protection provided by the mutual information suppression term with $\lambda=10$ has already exceeded the privacy level of $FSInfo=-0.5$, therefore impairing the accuracy slightly.

\begin{figure}[!t]
  \centering
  \includegraphics[width=\linewidth]{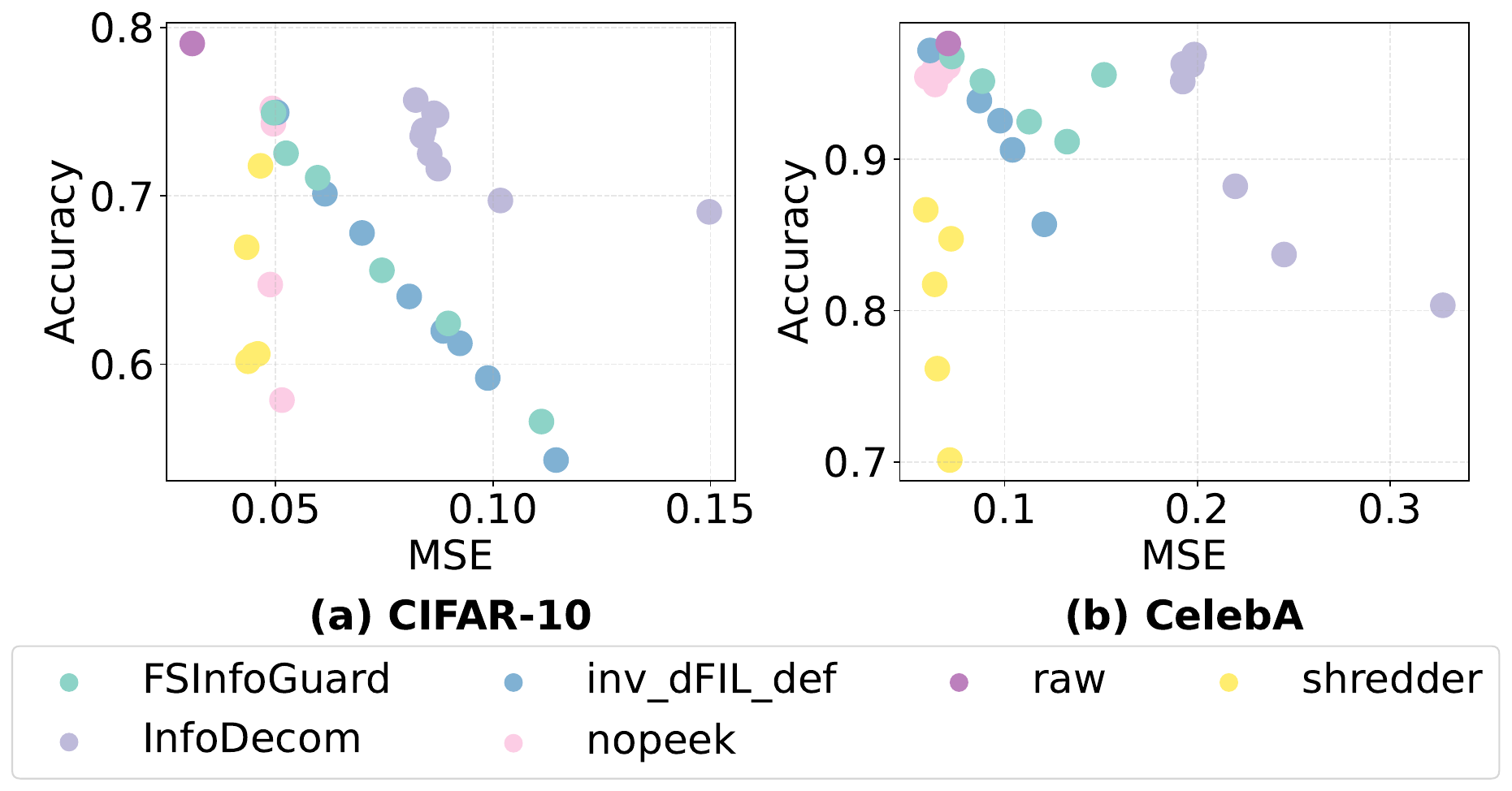}
  \caption{Model accuracy v.s. MSE on CIFAR-10 and CelebA against DRAs.}
  \label{fig:exp-pat-cifar10}
\end{figure}

\begin{figure}[!t]
  \centering
  \includegraphics[width=0.9\linewidth]{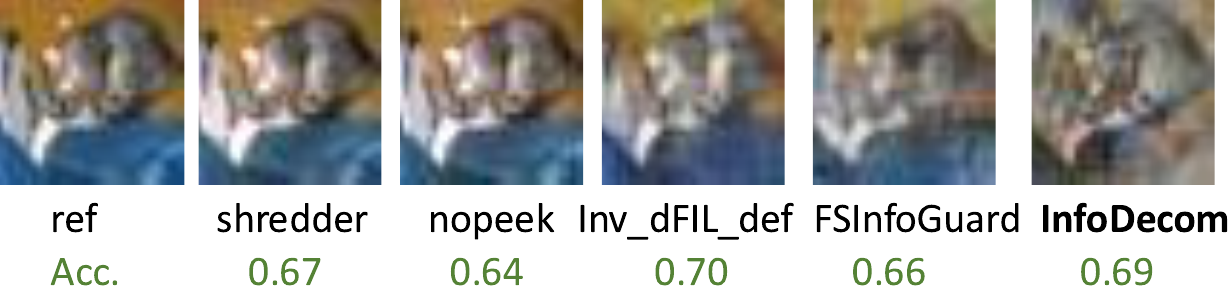}
  \caption{Images reconstructed by the DRAs along with the task accuracy on CIFAR-10 for different defenses.}
  \label{fig:exp-visual-cifar10}
\end{figure}
\subsection{Comparing Utility-Privacy Trade-off.}

We compare InfoDecom with four baselines in the utility-privacy plane, where the $x$-axis denotes MSE (privacy leakage) and the $y$-axis denotes task accuracy (utility), using both CelebA and CIFAR-10. Results of the undefended model (raw) are also included.
We vary the trade-off parameter of each to obtain different points on its curve. Specifically, for inv\_dFIL\_def we use $lb=\{0.01,0.08,0.2,0.4,1\}$; for FSInfoGuard, we use $FSInfo=\{-1.5,-1,-0.5,0.1,1\}$; for Nopeek, we use $\alpha=\{0.5,0.6,0.7,0.8,0.9\}$ (CelebA and CIFAR-10); for Shredder, we use $\text{coeff}=\{0.5,1,1.5,2,2.5\}$ (CelebA and CIFAR-10). For InfoDecom, we vary the $|X_h|$ (within $\{54,41\}$, the $\lambda$ (within [1,100]), and the $FSInfo$ (within $\{-0.5,-1,-1.5,-2\}$).
The results are shown in Figure~\ref{fig:exp-pat-cifar10}. \textbf{The InfoDecom achieves the best utility-privacy trade-off.} The reason is that it decomposes redundant information and retains only the necessary information to be protected.

To perceptually demonstrate the effectiveness of InfoDecom, we show the reconstructed images by DRAs on CIFAR-10 after applying the defenses in Figure~\ref{fig:exp-visual-cifar10}. The raw input is also given for reference. We see that, when the models have similar task accuracy, InfoDecom can better defend against privacy leakage.

\subsection{Ablation Studies}

\begin{table}[!t]
  \centering
  \begin{tabular}{ccc}
    \toprule
    Method & Acc. & MSE  \\
    \midrule
    \textbf{InfoDecom} & 0.7329  & 0.0843  \\
    \textbf{w/o} Vis. Rem. & 0.6273  & \textbf{0.0849} \\
    \textbf{w/o} $\mathcal{L}_{cl}$ & \textbf{0.7453}  &0.0835  \\
    \textbf{w/o} FSInfo & 0.7274  & 0.0826\\
    \bottomrule
  \end{tabular}
  \caption{CIFAR-10 on ResNet-18}
  \label{tab:ablation}
\end{table}
In this subsection, we demonstrate how each component in InfoDecom contributes to the overall improvement of the utility-privacy trade-off. We present the results in Table~\ref{tab:ablation}. The default value for $|X_h|$, $\lambda$, and $FSInfo$ is 54, 10, and -1. We can observe that, i) without visual information removal, the accuracy decreases. We find this is because noise with a larger scale is added to the smashed data to satisfy the required $FSInfo=-1$. ii) Without the mutual information suppression, although the model's predictive performance is largely preserved, the reduced regularization on smashed data may facilitate more accurate reconstructions by the attacker. iii) Without the FSInfo-guided noise perturbation, the level of privacy protection may fall short of the theoretically guaranteed bound.

\section{Conclusion}

We propose InfoDecom, a defense method designed to mitigate user data leakage in split inference (SI) for vision tasks by removing private information that is not essential for task performance. InfoDecom conducts a two-stage process to reduce redundant visual and mutual information. This is followed by a closed-form noise perturbation that ensures a theoretically guaranteed level of privacy protection. Experimental results show that InfoDecom achieves the best utility–privacy trade-off among the evaluated baselines. Although this work focuses on vision tasks, future research can explore the application of similar redundancy reduction strategies in natural language processing tasks.


\section{Acknowledgments}
This work is supported by Yangtze River Delta Science and Technology Innovation Community Joint Research Project (YDZX20233100004031).

\section{Regularization Terms of Defenses}
\begin{table}[!th]
    \caption{Regularization terms of existing methods. $x$ is the raw input data, $z$ is smashed data, $y$ is the true task label, ${\tilde{z}}$ is the perturbed smashed data, $\delta$ is the noise whose scale is $\sigma$, $\hat{x}$ is the reconstructed input, $u$ is the private attribute (e.g., the gender), $U$ is the uniform distribution, $x_c$ is the obfuscated input, $\Vert \cdot \Vert_p$ and $\Vert \cdot \Vert_q$ represents the p-norm and q-norm, and $DCOR$ is the distance correlation \cite{ICDM-20-nopeek}.} 
    \label{tab:rw-comparisons}

    \centering
    \setlength{\tabcolsep}{2pt} 
    \renewcommand{\arraystretch}{1.2} 
    \begin{tabular}{@{}ccc@{}}  
    \toprule
    Methods  & Goal~1 & Goal~2 \\
    \midrule
    Soteria & $- \| x - \hat{x} \|_p$ & $\text{s.t., } \| z - \tilde{z} \|_q \leq \epsilon$ \\
    ML-ARL&  $ KL(g(u|z) || U)$&  $ KL(p(y|x)||f(y|z))$ \\
    Cloak & $I(\mathbf{x}_c; \mathbf{u}) $& $ - I(\mathbf{x}_c; \mathbf{y})$ \\
    Nopeek & $  DCOR(\mathbf{X}, \mathbf{Z})$ &  cross-entropy \\
    Shredder & $ \frac{1}{\sigma^2(n)}$&  cross-entropy \\
    Inf2Guard &$ I(\mathbf{r} + \boldsymbol{\delta}; \mathbf{x}) - H(\boldsymbol{\delta})$&$ - I(\mathbf{r} + \boldsymbol{\delta}; y) + I(\mathbf{r}; y)$ \\
    TAPPFL &$ I(\mathbf{z}; u)$&$- I(\mathbf{x}; \mathbf{z} | u)$ \\
    ARPRL &$ I(\mathbf{z}; u)$&$- I(\mathbf{x}; \mathbf{z} | u)$\\
    DPFE &$ \quad I(\mathbf{z}; \mathbf{u})$& $-I(\mathbf{z}; \mathbf{y})$  \\
    InfoScissors &$\text{I}(\mathbf{z}; \mathbf{x})$& cross-entropy  \\

    \bottomrule
    \end{tabular}

\end{table}

\section{Detailed Experimental Implementation}
\subsection{Setups and Hyperparameters}



We use a server with Debian 6.1.119-1. InfoDecom has primarily focused on computer vision tasks so far. To this end, we adopt two widely adopted datasets: CIFAR-10 \cite{krizhevsky2009learning} and CelebA \cite{liu2015faceattributes}. CIFAR-10 is a simple and small dataset that allows rapid validation of methods, while the larger CelebA includes face images, a common privacy concern in the visual field, demonstrating the generalizability of the method on a complex dataset.

There are three main hyperparameters in InfoDecom. The number of retained coefficients $|X_h|\in \{1,\dots,64\}$, the weighting factor $\lambda \in [0,+\infty]$, and the $FSInfo\in[-\infty,+\infty]$. We chose hyperparameters that have generally performed well in multiple experiments.

\subsection{Attack Methods}
Data Reconstruction Attacks (DRAs) generally fall into two categories: neural network (NN)-based methods, which learn direct mappings from observable outputs to raw inputs, and maximum likelihood estimation (MLE)-based approaches that rely on optimization techniques such as stochastic gradient descent \cite{blackbox}. In federated learning or centralized machine learning systems, where only aggregated parameters or gradients are accessible, MLE-based methods are predominant, typically optimizing gradient matching objectives using Euclidean or cosine similarity \cite{zhang2023survey,geiping2020inverting,jin2021cafe}.

In contrast, SI scenarios favor NN-based attacks, as the one-to-one correspondence between raw inputs and smashed data offers abundant training pairs for inverse models, leading to superior performance over MLE-based techniques \cite{yin2023ginver,pasquini2021unleashing}. Recent advances in NN-based DRAs focus on mitigating performance degradation in practical settings by adapting the training set, loss functions, and decoder architectures—particularly when adversaries lack auxiliary datasets \cite{yin2023ginver} or must contend with defense mechanisms \cite{yang2022measuring}.

We adopt a permissive threat model where adversaries are allowed to use a standard yet effective inverse network \cite{blackbox} in our experiments. We use transposed convolutional layers \cite{dumoulin2016guide} to construct the decoder.

\subsection{Baselines}
\textbf{inv\_dFIL\_def} \cite{maeng2023bounding}. The Fisher information in the perturbed smashed data $z$ about the raw input $x$ is:
\begin{align}
    \mathcal{I}_z(x) &= \mathbb{E}[\nabla_x \log p(z;x) \nabla_x \log p(z;x)^{\top}]\\
    &= \frac{1}{\sigma^2}J^\top_{f_{\theta1}}J_{f_{\theta1}}, \label{eq:fil}
\end{align}
where the $\sigma$ is the scale of Gaussian noise, and $J_{f_{\theta1}}$ is the Jacobian of the bottom model $f_{\theta1}$ w.r.t. $x$.
Being viewed as the parameter estimation from $z$ to $x$, the data reconstruction attacker's estimation error can be bound by Fisher information as follows:
\begin{align}
\mathbb{E}[\Vert \hat{x}-x\Vert_2^2/d] \geq \frac{1}{dFIL}  \\
dFIL = Tr(\mathcal{I}_z(x))/d, \label{eq:crb}
\end{align}
where the dimension of input $x$ is $d$, $Tr(\cdot)$ denotes the matrix trace, $\hat{x}$ is the reconstructed input. From Equation~\ref{eq:fil}-\ref{eq:crb}, we can solve the Gaussian noise scale to achieve a certain dFIL:
\begin{equation}
    \sigma=\sqrt{\frac{Tr(J^\top_{f_{\theta1}}J_{f_{\theta1}})}{d \times dFIL}}.
\end{equation}
\cite{maeng2023bounding} calculate this noise scale in each training epoch and add the noise on smashed data. During split inference, the noise scale is similarly calculated and applied.

\textbf{FSInfoGuard} \cite{deng2025quantifying}.
FSInfoGuard has a similar procedure to inv\_dFIL\_def. However, the noise to be added is on a scale of:
\begin{equation}
    \frac{\det(J^\top_{f_{\theta1}}J_{f_{\theta1}})^{\frac{1}{2d}}}{e^{FSInfo}(2\pi e)^{\frac{1}{2}}},
\end{equation}
to achieve a certain privacy leakage, $FSInfo$.

\textbf{Nopeek} \cite{ICDM-20-nopeek}. Nopeek introduces such a regularization term to constrain the distance correlation between the input $x$ and smashed data $z$ during training:
\begin{equation}
    \mathcal{L} = \alpha DCOR(X;Z) + \text{cross-entropy},
\end{equation}
where the $DCOR(\cdot,\cdot)$ is the distance correlation and $\alpha$ is a scalar weighting factor.

\textbf{Shredder} \cite{mireshghallah2020shredder}. Shredder analyzes the mutual information $I(x;z)$ between $x$ and $z$ and defines a substitute version of it to regularize the bottom model. Shredder uses the reverse of the signal-to-noise ratio (1/SNR) as a proxy, where $SNR=\frac{\mathbb{E}[z^2]}{\sigma^2}$, with $\sigma$ the noise scale. The overall loss function during defensive training is:
\begin{equation}
    \mathcal{L} = \text{cross-entropy} - \alpha |\delta|,
\end{equation}
where the $|\delta|$ is the norm of the noise.

\section{Computational Efficiency}
We examine the computational efficiency of InfoDecom in this section. InfoDecom introduces overheads for achieving better privacy-utility trade-off, which is a common cost incurred by Jacobin-based methods (e.g., FSInfoGuard, inv\_dFIL\_def). Methods that avoid Jacobians (e.g., Nopeek, Shredder) are more efficient but may typically suffer worse privacy-utility tradeoffs. We employ some techniques \cite{deng2025quantifying} to reduce the overheads: input downsampling, diagonal approximation of the Fisher Information Matrix, and parallel processing. For an input dimension of d and the number of client model parameters P, the complexity is approximately O(dP). Also, we present the run-time measurements in Table~\ref{tab:inference-time}.

The delay of InfoDecom is comparable to that of other Jacobian-based methods. The absolute additional delay (\~6 ms) introduced by InfoDecom is acceptable for typical split learning applications such as financial risk detection and personalized recommendation. These scenarios prioritize data privacy and security over ultra-low latency, and a few milliseconds of extra delay do not affect normal functionality or user experience. We also plan to explore more efficient Jacobian approximation strategies in future work to further reduce client-side overhead.


\begin{table}
    \centering
    \begin{tabular}{cccccc}
    \toprule
    \makecell{Forward} & \makecell{No-\\peek} & \makecell{Shred-\\der} & \makecell{FSInfo-\\Guard} & \makecell{inv\_dFIL-\\\_def} & \makecell{Info\\Decom} \\
    \midrule
    0.26 & 0.94 & 0.45 & 4.52 & 4.90 & 6.64 \\
    \bottomrule
    \end{tabular}
    \caption{Inference time (ms) for one sample on ResNet-18 and CelebA. Forward means the feed-forward time of the bottom model.}
    \label{tab:inference-time}
\end{table}

\section{More than Images}
The proposed InfoDecom consists of three components: visual information removal (VIR), mutual information suppression (MIS), and noise perturbation (NP). While VIR currently targets image data, MIS and NP are modality-agnostic.

More importantly, the ''decompose-then-noise'' framework and the insight that redundant information harms the privacy–utility tradeoff are general and form our main contribution. Future work can replace VIR with broader redundancy-removal modules (e.g., for text or multimodal data) to extend InfoDecom beyond images.

\bibliography{aaai2026}
\end{document}